\documentclass[11pt,tightenlines,superscriptaddress,floatfixolumn,english,aps,notitlepage,nofootinbib]{revtex4-1}
\usepackage{amsmath}
\usepackage{esint}   
\usepackage{graphicx}
\usepackage{hyperref}
\usepackage{color}
\usepackage{xcolor}
\usepackage{enumerate}
\usepackage[utf8]{inputenc} %%% needed  for some of the references
\advance\voffset -0.2in

\newcommand{\be}{\begin{eqnarray}}
\newcommand{\ee}{\end{eqnarray}}
\newcommand{\non}{\nonumber\\}

\newcommand{\ave}[1]{\left\langle #1 \right\rangle}
\newcommand{\avet}[1]{\langle #1 \rangle}  % in text

\newcommand{\gev}{{\rm \, GeV}}

\definecolor{vkcolor2}{HTML}{336600}

\newcommand{\overbar}[1]{\mkern 1.5mu\overline{\mkern-1.5mu#1\mkern-1.5mu}\mkern 1.5mu}
\newcommand{\nb}{\overbar{N}}
\newcommand{\dc}[1]{\overbar{C}_{#1}}
\begin{document}

%\preprint{}
\title{
Large proton cumulants from the  superposition of ordinary multiplicity distributions
}

\author{Adam Bzdak}
\email{bzdak@fis.agh.edu.pl}
\affiliation{AGH University of Science and Technology,
Faculty of Physics and Applied Computer Science,
30-059 Krak\'ow, Poland}

\author{Volker Koch}
\email{vkoch@lbl.gov}
\affiliation{Nuclear Science Division, 
Lawrence Berkeley National Laboratory, 
Berkeley, CA, 94720, USA}

\author{Dmytro Oliinychenko}
\email{DOliinychenko@lbl.gov}
\affiliation{Nuclear Science Division, 
Lawrence Berkeley National Laboratory, 
Berkeley, CA, 94720, USA}

\author{Jan Steinheimer}
\email{steinheimer@fias.uni-frankfurt.de}
\affiliation{Frankfurt Institute for Advanced Studies, Ruth-Moufang-Str. 1, D-60438 Frankfurt am Main, Germany}

\begin{abstract}
We construct a multiplicity distribution characterized by
large factorial cumulants (integrated correlation functions) from a simple combination 
of two ordinary multiplicity
distributions characterized by small factorial cumulants. We find that such a model, which could
be interpreted as representing two event classes, reproduces the preliminary data for the proton
cumulants measured by the STAR collaboration at $7.7$ GeV very well. This model then predicts very large values
for the fifth and sixth order factorial cumulants, which can be tested in experiment.
\end{abstract}

\maketitle

\section{Introduction}
The study of fluctuations and correlations of conserved charges has become the focus of attention in
the search of the QCD phase transition and its conjectured critical point
\cite{Stephanov:2004wx,Stephanov:2008qz,Skokov:2010uh,Stephanov:2011pb,Luo:2011rg,Luo:2017faz,Herold:2016uvv,Zhou:2012ay,Wang:2012jr,Karsch:2011gg,Schaefer:2011ex,Chen:2011am,Fu:2009wy,Cheng:2008zh,Jeon:2000wg,Gazdzicki:2003bb,Gorenstein:2003hk,Gazdzicki:2003bb,Koch:2005vg,Alt:2007jq,Alt:2008ab}. While
at vanishing chemical potential lattice QCD calculations have established that the transition occurs
as an analytic crossover \cite{Aoki:2006we,Borsanyi:2010cj,Bazavov:2011nk}, no real constraints on
the phase transition at large chemical potential are currently possible. Therefore one mainly relies
on experimental studies to identify the order of the QCD phase transition. If indeed a first order
phase transition occurs at large baryon chemical potentials and intermediate temperatures, also an
associated endpoint must exist at which the transition becomes second order.  It is well known that
first and second order phase transitions give rise to many interesting phenomena, especially related
to fluctuations and long range correlations \cite{Jeon:2000wg,Koch:2008ia}. In macroscopic systems
that slowly approach a second order phase transition the correlation length diverges and similarly
systems that approach a first order phase transition will undergo nucleation, leading to droplet
formation, or spinodal decomposition
\cite{Chomaz:2003dz,Randrup:2003mu,Sasaki:2007db,Randrup:2009gp,Nahrgang:2011mg,Mishustin:1998eq,Steinheimer:2012gc}. One
observable of particular interest are higher order cumulants of the proton number distribution. It has been
shown \cite{Stephanov:2008qz} that, close to the critical point, higher orders of the cumulants
diverge with ever increasing powers of the correlations length.

The goal of several experimental programs at RHIC, SPS, NICA, J-PARC and GSI/FAIR facilities is to
identify observables which would show the increase of particle correlations and fluctuations due to
a phase transition. To do so, heavy nuclei are brought to collision at relativistic energies. The
system created in such collisions may have densities several times that of normal nuclear 
saturation density and temperatures of more than 200 MeV \cite{Damjanovic:2009zz,Specht:2010xu},
a finding supported by microscopic transport simulations as well as fluid dynamical simulations 
(see e.g. \cite{Rapp:2016xzw,Endres:2015egk,Steinheimer:2009nn}) at the energies discussed.

In addition, by varying the beam energy of
the collision one is able to change the density and temperature of the created system, which allows to 
'scan' the phase diagram of QCD and hopefully locate the onset of phase transition signals.

The systems created in these heavy ion collisions are very small (a few fm in size) rapidly
expanding (the expansion velocity is close to the speed of light) and short lived (the system
decouples after 10-20 fm/c). These features affect possible signals of a
phase transition. For example, there may not be sufficient time for the  correlation length to grow significantly
near the critical point \cite{Berdnikov:1999ph} or for nucleation, 
which is also a comparatively slow process, to occur. On the other hand
phenomena like spinodal decomposition, i.e., the rapid phase separation due to (mechanical) instabilities at the
phase transition are much faster due to the exponential growth of fluctuations in the mechanically unstable region, 
and thus can occur resulting in an increase of density fluctuations
\cite{Steinheimer:2012gc,Herold:2013bi}. Further complications
arise from the fact  that the number of charge, strangeness or baryon number carrying particles is relatively small, especially for the baryon number. As a consequence, effects of the global
conservation of the various conserved charges cannot be neglected \cite{Begun:2004gs,Schuster:2009jv,Bzdak:2012an}. In addition, as the system
rapidly drops out of equilibrium, other effects like resonance decays, thermal smearing may blur the
signal. Additionally, experimental acceptance and efficiency corrections need to be taken into
account \cite{Bzdak:2012ab,Bzdak:2016qdc,Kitazawa:2016awu,He:2018mri,Feckova:2015qza,Begun:2004gs,Bzdak:2012an,
Gorenstein:2008et,Gorenstein:2011vq,Sangaline:2015bma,Tarnowsky:2012vu,
Xu:2014jsa,Adamczyk:2014fia,Adamczyk:2013dal}.

Published data on the net-proton number fluctuations, which, with reasonable model assumptions, can
be related to the net-baryon fluctuations \cite{Kitazawa:2011wh,Kitazawa:2012at}, only exist from
the STAR collaboration \cite{Adamczyk:2013dal,Aggarwal:2010wy} and for a limited acceptance. The published data
($|y|<0.5$ and $0.4<p_t<0.8$ GeV) are consistent with uncorrelated proton production and the trivial
correlations from global baryon conservation \cite{Bzdak:2012an}.  On the other hand preliminary
data from the STAR collaboration with a larger acceptance ($0.4<p_t<2$ GeV)
\cite{Luo:2015ewa,Luo:2015doi} show a significant deviation from uncorrelated proton emission,
i.e., following Poisson statistics, for collision energies 
$\sqrt{s_{NN}}\leq 11\gev$. The preliminary data consistently show an increase of the
fourth order cumulant and a decrease of the third order cumulant with respect to uncorrelated
production, in particular at the lowest beam energies of $\sqrt{s_{NN}}=7.7 \gev $ and $\sqrt{s_{NN}}=11 \gev$.
Presently, higher order net-proton cumulants are also being studied in other experiments, such as HADES
\cite{Lorenz:2017hjp} and NA61 \cite{Davis:2017mzd,Gazdzicki:2017zrq}.

The experiments provide the measured cumulants of the net-proton
number distributions and not the actual multi-particle correlation
functions. However, the integrated n-particle correlation functions (factorial cumulants) can be
extracted from the measured cumulants \cite{Bzdak:2016sxg} and they
indeed show an interesting beam energy dependence. In particular, the
integrated four particle correlations at the lowest beam energy
accessible to STAR, $\sqrt{s_{NN}}=7.7\gev$ are very large, about three
orders of magnitude larger than a basic Glauber model (incorporating the number of 
wounded nucleons \cite{Bialas:1976ed} fluctuations) combined
with baryon number conservation would predict \cite{Bzdak:2016jxo}. The challenge now is to
unambiguously connect the measured correlations to physical effects from a critical point or first
order phase transition. 

In this paper we will investigate how one can construct a proton multiplicity distribution,
consistent with the unexpectedly large, as compared to expectations from conventional models
\cite{Bzdak:2016jxo,He:2017zpg}, factorial cumulants extracted 
\cite{Bzdak:2016sxg} from the recent preliminary STAR measurements at $\sqrt{s_{NN}}=7.7\gev$.
To this end we will focus on the case where the proton
distribution function results from a superposition of two independent distributions, or sources of
protons. In particular we want to explore if it is possible to construct such a superposition for
the case where individual distributions are both characterized by {\em small} factorial cumulants.
In principle there are two distinct ways to construct such a superposition of two
sources/distribution: In the first scenario, one
assumes that in each event the protons arise from both sources at the same time and thus the total proton number is
simply the sum of the protons drawn from the two distributions. As we shall discuss in the paper,
in this case the factorial cumulants of the combined distribution are of the same order of magnitude
as those of the two individual distributions. In the second scenario, one assumes that  the superposition of two
independent distributions is such that in a given event the proton multiplicity is drawn exclusively,
with a certain probability, from either one of the two distributions. This case corresponds
to having two distinct event classes, and as we shall show, the factorial cumulants of the combined
distribution can be very large even if the factorial cumulants of the individual distributions are small or even
vanish as it would be the case if we were using Poissonians. It is this second scenario which we
will concentrate on in this paper. In particular we will discuss the extracted \cite{Bzdak:2016sxg} factorial cumulants
from the STAR experiment in the context of such multiplicity distributions. Possible interpretations in terms of phase transition physics and 'non-physics' background will be given. Furthermore we will propose further experimental studies which will help to better understand the origin of experimentally measured large correlations.

\section{Two events classes}

Let us consider the situation where we have two different types (or
classes)  of events, denoted by (a) and (b). Let us denote the
probability that an event belongs to class (a) by $(1-\alpha)$ and to
class (b) by $\alpha$ with $\alpha \leq 1$. In this case the probability
to find $N$ particles or protons is given by
\begin{equation}
P(N)=(1-\alpha )P_{(a)}(N)+\alpha P_{(b)}(N),
\label{eq:two_component}
\end{equation}%
where $P_{(a)}(N)$ and $P_{(b)}(N)$ are multiplicity distributions governing the event classes (a)
and (b) respectively. As we shall show, the combined distribution, Eq.~\eqref{eq:two_component}, can
exhibit very large factorial cumulants (integrated correlation functions) even if neither $P_{(a)}$
nor $P_{{(b)}}$ exhibit any correlations, as would be the case if they were Poissonian.  
Such a
situation can arise for example if in a heavy ion experiment the centrality selection, for whatever
reasons, mixes central events with very peripheral ones.

In order to calculate the factorial cumulants it
is best to start with the generating function:
\begin{eqnarray}
H(z) &=&\sum P(N)z^{N}  \notag \\
&=&(1-\alpha )H_{(a)}(z)+\alpha H_{(b)}(z).
\end{eqnarray}%
where $H_{(a),(b)}$ is the generating function for $P_{(a),(b)}$. The factorial cumulant generating function is then given by%
\begin{eqnarray}
G(z) &=&\ln \left[ H(z)\right]  \notag \\
&=&\ln \left[(1-\alpha )H_{(a)}(z)+\alpha H_{(b)}(z)\right]  \notag \\
&=&\ln \left[ (1-\alpha )e^{\ln [H_{(a)}(z)]}+\alpha e^{\ln [H_{(b)}(z)]}%
\right]  \notag \\
&=&G_{(a)}(z) + \ln \left[ 1-\alpha + \alpha e^{G_{(b)}(z)-G_{(a)}(z)}\right] ,
%%&=&\ln \left[ (1-\alpha )e^{G_{(a)}(z)}+\alpha e^{G_{(b)}(z)}\right] ,
\label{eq:generating_func}
\end{eqnarray}%
where $G_{(a),(b)}(z)= \ln[H_{(a),(b)}(z)]$. The  factorial cumulants read\footnote{Following, e.g., Ref. \cite{Bzdak:2016sxg} we denote 
the factorial cumulants (integrated multi-particle correlation functions) by $C_n$ and the cumulants
by $K_n$. This notation should not be confused by the one
adapted by the STAR collaboration, which denote the cumulants by $C_{n}$.}
\begin{equation}
C_{k}=\left. \frac{d^{k}}{dz^{k}}G(z)\right| _{z=1},
\end{equation}%
and analogously for $C_{k}^{(a)} = \frac{d^{k}}{dz^{k}}G_{(a)}(z)|_{z=1}$ 
and $C_{k}^{(b)} = \frac{d^{k}}{dz^{k}}G_{(b)}(z)|_{z=1}$.

Given the distribution Eq.~\eqref{eq:two_component}, the 
mean number of protons is 
\begin{equation}
\left\langle N\right\rangle =(1-\alpha )\left\langle N_{(a)}\right\rangle
+\alpha \left\langle N_{(b)}\right\rangle ,
\end{equation}%
with $\ave{N_{(a),(b)}}=\sum_{N} N P_{(a),(b)}(N)$ is the average particle
number for distributions $P_{(a)}(N)$ and $P_{(b)}(N)$, respectively. 
To simplify the notation we further introduce%
\begin{eqnarray}
\nb  &=&\left\langle N_{(a)}\right\rangle -\left\langle
N_{(b)}\right\rangle ,  \notag \\
\dc{n} &=&C_{n}^{(a)}-C_{n}^{(b)},
\end{eqnarray}%
and performing straightforward calculations we obtain\footnote{The formulas for higher orders 
are given in the Appendix \ref{sec:app_C5_and_C6}.},\footnote{We note that the terms involving
$\alpha$ such as  $\alpha$, $\alpha(1-\alpha)$, $\alpha(1-\alpha)(1-2\alpha)$ etc. appearing in
Eq~\eqref{eq:C234-full} are simply the cumulants of the Bernoulli distribution. This
becomes apparent from Eq.~\eqref{eq:generating_func}, since upon replacing
$G_{(b)}(z)-G_{(a)}(z)=t(z)$ the second term represents the
cumulant generation function of the Bernoulli distribution.} 
\begin{eqnarray}
C_{2} = C_{2}^{(a)}&-&\alpha  \left\{\dc{2}-(1-\alpha ) \nb^2\right\} ,
\non
C_{3} = C_{3}^{(a)}&-&\alpha  \left\{ \dc{3}+(1-\alpha ) \left[ (1-2
         \alpha ) \nb^3 -3 \nb \dc{2} \right] \right\} ,
\non
C_{4} = C_{4}^{(a)}&-&\alpha  \left\{ \dc{4}-(1-\alpha ) \left[
         \left(1-6 \alpha +6 \alpha ^2\right) \nb^4 \right. \right. \non
        &  - & \left. \left. 6  (1-2 \alpha ) \nb^2 \dc{2} + 4 \nb \dc{3} 
         + 3 (\dc{2})^{2} \right] \right\} .
    \label{eq:C234-full}
\end{eqnarray}
So far the results are general and apply for an arbitrary choice of distributions $P_{(a)}$ and $P_{(b)}$.

Our goal here is to obtain large factorial cumulants $C_n$ from {\it ordinary} multiplicity distributions
characterized by small factorial cumulants, namely $C_{n}^{(a)} \ll C_n$ and $C_{n}^{(b)} \ll
C_n$. This is motivated by a surprisingly large three- and four-proton factorial cumulants, $C_{3}$
and $C_{4}$, measured in central Au+Au collisions at $\sqrt{s_{NN}} = 7.7$ GeV, which are much larger
than simple expectations from baryon conservation or $N_{\rm part}$ fluctuation. The ultimate case
where this holds is when the two classes are governed by Poisson distributions,
where  $C_{n}^{(a)} = C_{n}^{(b)} = 0$. In this case the factorial
cumulants reduce to
\begin{eqnarray}
C_{2} &=& \alpha (1-\alpha) \nb^{2}
\approx \alpha \nb^{2} , \non
C_{3} &=& -\alpha (1-\alpha ) (1-2\alpha) \nb^{3}
 \approx -\alpha \nb^{3} , \non
C_{4} &=& \alpha (1-\alpha) (1-6\alpha +6\alpha^{2}) \nb^{4}
\approx \alpha \nb^{4},
\label{eq:poisson_limit} 
\end{eqnarray}%
where in the last step we assumed $\alpha \ll 1$. In general, for small $\alpha$ we have ($n > 1$)  
\begin{equation}
C_{n} \approx (-1)^{n}\alpha\nb^{n}, \quad \alpha \ll 1, \quad n > 1
\label{eq:cn-poiss-small-alpha}
\end{equation}
and the higher order factorial cumulants can assume very large values even for small values of $\alpha$. 

In general, if the measured $C_n \gg C_{n}^{(a)}$ and $C_n \gg C_{n}^{(b)}$, we obtain
Eqs. (\ref{eq:poisson_limit}) and (\ref{eq:cn-poiss-small-alpha}). In this case, $C_n$ is
independent of the details of $P_{(a)}(N)$ and $P_{(b)}(N)$, and the signal is driven almost
exclusively by the superposition between the two distributions. This results in the relation between
factorial cumulants of adjacent order 
\begin{equation}
\frac{C_{n+1}}{C_n} \approx -\nb,
\label{eq:ratio}
\end{equation}
% \begin{equation}
% \frac{C_{n+1}}{C_n} = -\nb + \mathcal{O}(\alpha \nb) \approx -\nb,
% \label{eq:ratio}
% \end{equation}
for $\alpha \ll 1$ with the first correction being $\mathcal{O}(\alpha \nb)$. Also in this limit, the above ratio does not depend on $\alpha$.  At $7.7$ GeV, $C_{4}/C_{3} \sim -17$, see Ref. \cite{Bzdak:2016sxg}, (with admittedly large error bars). Our approach clearly predicts the ratio of higher order cumulants, as well as the fact that their signs are alternating, see Eq. (\ref{eq:cn-poiss-small-alpha}), which can be tested by future experimental data on $C_{5}/C_{4}$ and $C_{6}/C_{5}$.\footnote{Using $C_{2}$ in this analysis is not advised since, e.g., at $7.7$ GeV the measured $C_{2}$ is consistent with an ordinary background and the condition $C_n \gg C_{n}^{(a)}$ and $C_n \gg C_{n}^{(b)}$ is not satisfied.} To summarize, if indeed the large factorial cumulants observed at $7.7$ GeV originate from the superposition of two event classes with $\alpha \ll 1$, we expect
\begin{equation}
\frac{C_{6}}{C_5} \approx \frac{C_{5}}{C_4} \approx \frac{C_{4}}{C_3} = -17 \pm 6 ,
\label{eq:ratio-6543}
\end{equation}
where the uncertainty is based on $C_4 = 170 \pm 45$ and $C_3 = -10 \pm 2.5$ \cite{Bzdak:2016sxg}.

In \cite{Bzdak:2016sxg,Bzdak:2017ltv} it was shown that the integrated correlation functions
(factorial cumulants) obtained from the STAR measurements are consistent with rapidity and
transverse momentum independent normalized multi-particle correlation functions. In other 
words, the STAR data are consistent with a very large correlation length in rapidity and
transverse momentum.  This in turn implies that the factorial cumulants scale with the mean
proton number like $C_{n}\sim \avet{N}^{n}$. As seen from Eqs. (\ref{eq:poisson_limit}) and
(\ref{eq:cn-poiss-small-alpha}), valid if $C_n \gg C_{n}^{(a)}$ and $C_n \gg C_{n}^{(b)}$, this is
naturally explained in our superposition model. Indeed, if we were really dealing with two event
classes, the relative weight of the two distributions, $\alpha$, would be independent of the size of
the rapidity window, whereas the mean number of particles from event types (a) and (b) would roughly
scale with the rapidity window $\avet{N_{(a)}} \sim \Delta Y$ and $\avet{N_{(b)}} \sim \Delta
Y$. Consequently $\avet{N}\sim \Delta Y$ and $\nb \sim \Delta Y$. As a result,
$\nb \sim \Delta Y \sim \avet{N}$. Of course, this observation constitutes no proof for the
existence of two event classes in the STAR data, as there may very well be other distributions with
a similar scaling, but it certainly is a nice consistency check.

We note that our model is not well suited for multiplicity distributions characterized by small
$C_n$. In this case the details of $C_{n}^{(a)}$ and $C_{n}^{(b)}$ are crucial and we loose any
predictive power. However, if $C_n$ is small, there is most likely no need to introduce two event classes and the signal may very well be explained by an {\it ordinary} background. This is, e.g., the case for $\sqrt{s_{NN}}= 19$ GeV collisions, where the measured $C_3$ and $C_4$ are close to zero and, in fact, are consistent (withing larger error bars) with simple baryon conservation with possible contribution from $N_{\rm part}$ fluctuation.     

The main finding of this Section is that our model of two event classes indeed leads to large values
for the higher order factorial cumulants and that we predict a straightforward relation between
them, Eqs.~\eqref{eq:ratio} and \eqref{eq:ratio-6543}, which can be tested in experiment. Next let
us turn to a somewhat more refined analysis of the STAR data at $\sqrt{s_{NN}}=7.7\gev$.

\subsection{Data analysis} 
From the analysis \cite{Bzdak:2016sxg} of the preliminary STAR data
\cite{Luo:2015ewa} we know that for central collisions at
$\sqrt{s_{NN}}=7.7\gev $ the factorial
cumulants of the proton multiplicity distribution up to fourth order are 
\begin{eqnarray}
\text{STAR}\text{: } &&\langle N \rangle \approx 40,\,  C_{2}\approx -2,\,  C_{3}\approx -10,\,
                        C_{4}\approx 170. \nonumber %\notag
\label{eq:STAR_fact_cumulants}
\end{eqnarray}%
and, as already pointed out, $C_3$ and $C_4$ are much larger than
expectations from an ordinary background, such as baryon conservation and $N_{\rm part}$ fluctuation
\cite{Bzdak:2016jxo}.  

Thus we are in the situation discussed in the previous section, where the superposition of two ordinary multiplicity distributions, Eq. (\ref{eq:two_component}), can easily
generate large factorial cumulants which are independent on the specific choice for $P_{(a)}(N)$ and
$P_{(b)}(N)$, provided $C^{(a)}_{n}$ and $C^{(b)}_{n}$ are much smaller then the measured $C_{n}$, see
Eqs. (\ref{eq:poisson_limit}) and (\ref{eq:cn-poiss-small-alpha}). The simplest choice is to take
Poisson distributions for both $P_{(a)}$ and $P_{(b)}$. The next refinement is to use a binomial
distribution for $P_{(a)}$ in order to capture the effect of baryon number conservation
\cite{Bzdak:2016jxo}. This actually results in $C_{2}<0$, as seen in the data.

Consequently, we take $P_{a}(N)$ as binomial,
\begin{equation}
	P_{a}(N)=\frac{B!}{N!(B-N)!}p^{N}(1-p)^{B-N} ,
\end{equation}
with $B=350$, which properly captures baryon number
conservation, and $P_{b}(N)$ as Poisson.\footnote{We could also chose binomial here but this is
rather irrelevant for our results. For example, $C_2$ depends on $C^{(b)}_{2}$ through
$\alpha \dc{2} $ which is expected to be much smaller than $C^{(a)}_{2}$. An actual fit to two
binomials results in $C_{2}=-4.03$ which, given the uncertainty of the contribution due to participant
fluctuations \cite{Bzdak:2016jxo}, is in equally good agreement with the STAR data. At the same time the predictions for
$C_{5}$ and $C_{6}$ are within 3\% of those using just one binomial.} In this case the relevant 
factorial cumulants are given by
\begin{eqnarray}
C_{2}^{(a)} &=&-p^{2}B,\quad C_{3}^{(a)}=2p^{3}B,\quad C_{4}^{(a)}=-6p^{4}B , 
\notag \\
C_{5}^{(a)} &=&24p^{5}B,\quad C_{6}^{(a)}=-120p^{6}B ,
\end{eqnarray}%
with $\langle N_{(a)}\rangle =pB$. Obviously $C^{(b)}_{n}=0$ and $\dc{n}=C^{(a)}_{n}$. 

Using Eqs. (\ref{eq:C234-full}) we fit the mean number of protons as well as the third and the fourth order factorial
cumulants resulting in
\begin{eqnarray}
\alpha \approx 0.0033, \quad \nb\approx 14.7,\quad p \approx 0.114, %\notag
\end{eqnarray}
which also gives $\langle N_{(a)}\rangle \approx 40$ and $\langle N_{(b)}\rangle \approx 25.3$.
We note that indeed $\alpha\ll 1$ as assumed in Eqs. (\ref{eq:cn-poiss-small-alpha}), (\ref{eq:ratio}) and (\ref{eq:ratio-6543}). 

\begin{figure*}[t]
\begin{center}
\includegraphics[scale=0.38]{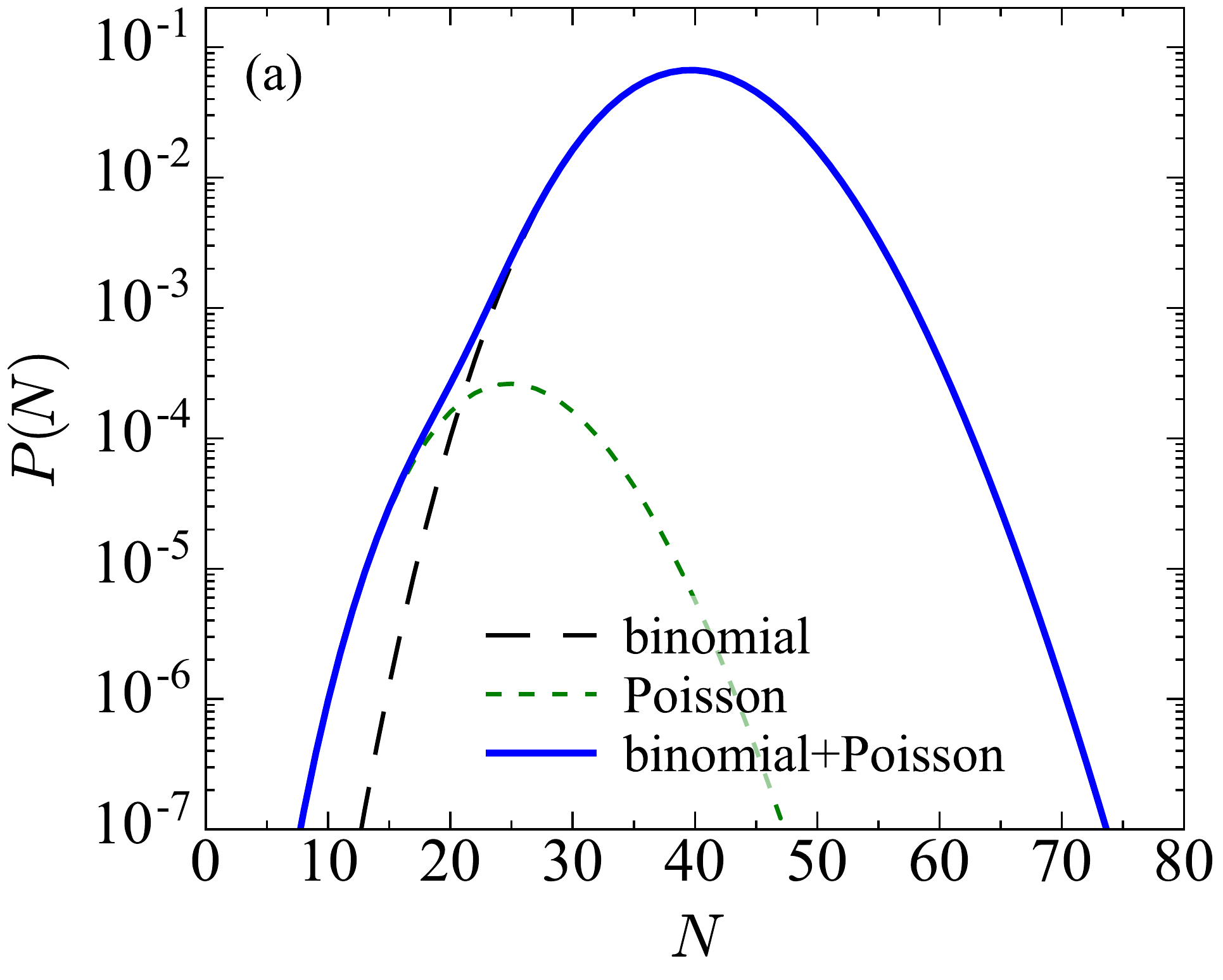}\hspace{10mm} %
\includegraphics[scale=0.38]{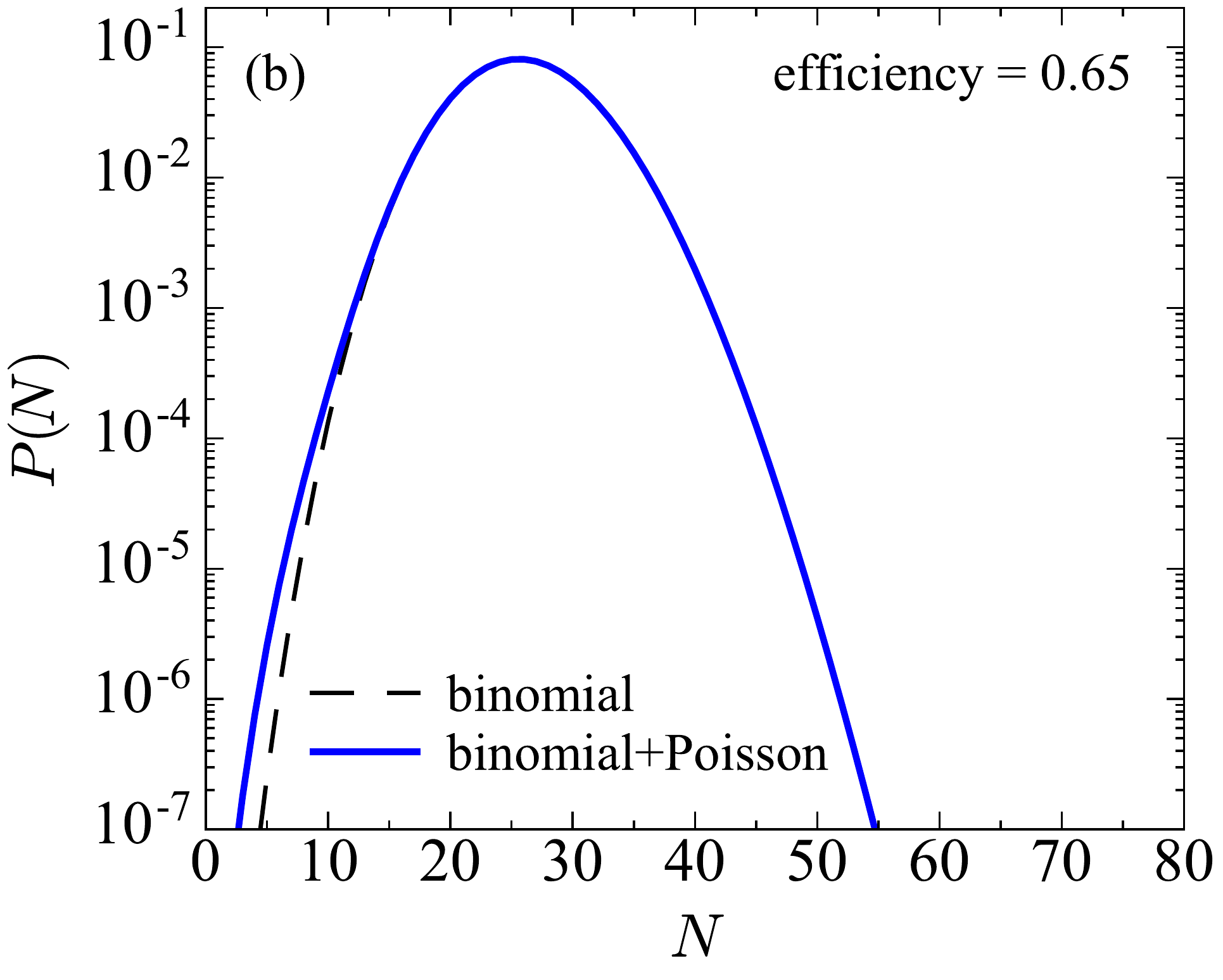}
\end{center}
%\par
%\vspace{-5mm}
\caption{The multiplicity distribution $P(N)$ at $\sqrt{s_{NN}}=7.7 \gev$  in the two component model given by Eq. (\ref{eq:two_component}) constructed with (a) efficiency unfolded values for $\langle N \rangle$, $C_3$ and $C_4$ and (b) with imposed efficiency of $0.65$. }
\label{fig:1}
\end{figure*}

Given the fit, we can also predict the factorial cumulants, $C_2$,
$C_5$, $C_6$ and we obtain:\footnote{Taking $C_4 = 130\,(210)$, 
being consistent with the preliminary STAR data \cite{Bzdak:2016sxg}, we obtain 
$\alpha\approx 0.0078 \, (0.0017)$, $\nb \approx 10.92 \,(18.43)$, 
$p\approx 0.115 \, (0.114)$, and $C_2 \approx -3.64 \, (-3.99)$, 
$C_5 \approx -1546 \, (-4030)$, $C_6 \approx 17970 \, (77229)$.
Also $K_5/K_2 = -14 \, (-61)$ and $K_6/K_2 = 62 \, (818)$. 
For larger $C_4$, the value of $\alpha$ gets smaller but $\nb$ gets larger, 
which is more effective in increasing the value of $C_4$, see Eq. (\ref{eq:poisson_limit}). }
\begin{align}
C_2 &\approx -3.85 , \non
C_5 &\approx -2645 , \non
C_6 &\approx 40900 ,
  \label{eq:C56_predict}
\end{align}
which corresponds to the following values for the cumulant ratios\footnote{$K_{2} = \langle N\rangle + C_{2}$, $K_{5}=\langle N\rangle + 15
      C_{2} + 25 C_{3} + 10 C_{4} + C_{5}$ and $K_{6}= \langle N\rangle + 31 C_{2} + 90 C_{3} + 65
      C_{4} + 15 C_{5} + C_{6}$.} 
\begin{align}
K_5/K_2 &\approx -34 , \non
K_6/K_2 &\approx 312 .
  \label{eq:K_predict}
\end{align}
It is worth pointing out that $C_6/C_5 \approx C_5/C_4 \approx C_4/C_3$ in agreement with the discussion presented in the previous Section. 
We note that the resulting $C_2 \approx -3.85$ is
slightly more negative than the data. However, as shown, e.g., in
\cite{Bzdak:2016jxo}, the second order factorial cumulant receives
a sizable positive contribution from participant fluctuations $\Delta
C_{2}\simeq 2-3 $ whereas the correction to $C_{3}$ and $C_{4}$ are
small. In any case correcting data for the fluctuations 
of $N_{\rm part}$ should be done very carefully to avoid model dependencies. 
In view of the sizable errors in the preliminary STAR data we
consider the present fit as satisfactory.

The resulting probability distribution for the proton number, $P(N)$,
Eq.~\eqref{eq:two_component}, is shown in the left panel of
Fig.~\ref{fig:1}.\footnote{Since we extract the multiplicity distribution from bin width corrected cumulants, our result
corresponds to an appropriately bin width corrected multiplicity distribution.} Even though the component centered at $N\sim 25$ has
a very small probability $\alpha\sim 0.3\%$ it
gives rise to a shoulder at low $N$ which should be visible
in the multiplicity distribution. However, this would require an
unfolding of the measured distribution \cite{Bzdak:2016qdc} in order to remove
the effect of a finite detection efficiency. Assuming a binomial model
for the efficiency with a constant detection probability of  
$\epsilon = 0.65$, which roughly corresponds to that of the
STAR measurement, the observed multiplicity distribution of the two
component model is shown in the right panel of Fig.~\ref{fig:1}. In
this case the small component $\sim
\alpha$ is barely visible. This observation is consistent with the
fact that the efficiency {\em uncorrected} cumulants measured by STAR
are more or less consistent with a Poisson (or binomial to be more precise) expectation.    

Another complication arises from the fact that the proton distribution functions are only measured
in rather broad centrality bins. Directly correcting the distribution functions for this centrality
bin-width effect is not straight forward. One should also note, that we only considered the most
central data, as only here one observes significantly large correlations, i.e.,
$C_{n}^{(a)} \ll C_n$. For more peripheral centrality bins the situation is less clear,
as the correlations, $C_{n}$, become smaller, which does not allow for a clear distinction between a
single ordinary distribution and the superposition of two distributions.

\begin{figure*}[t]
\begin{center}
\includegraphics[scale=0.38]{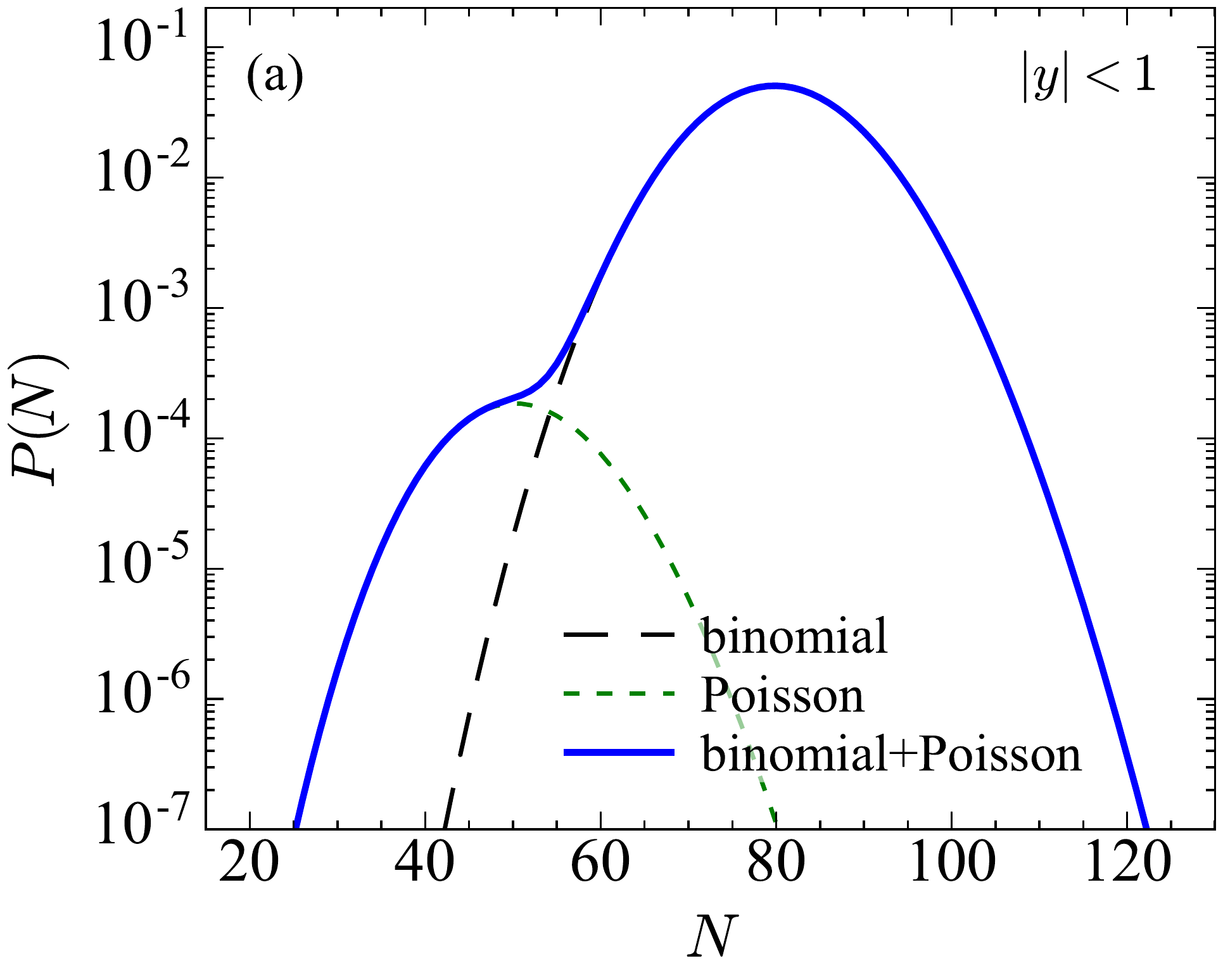} \hspace{10mm}
\includegraphics[scale=0.38]{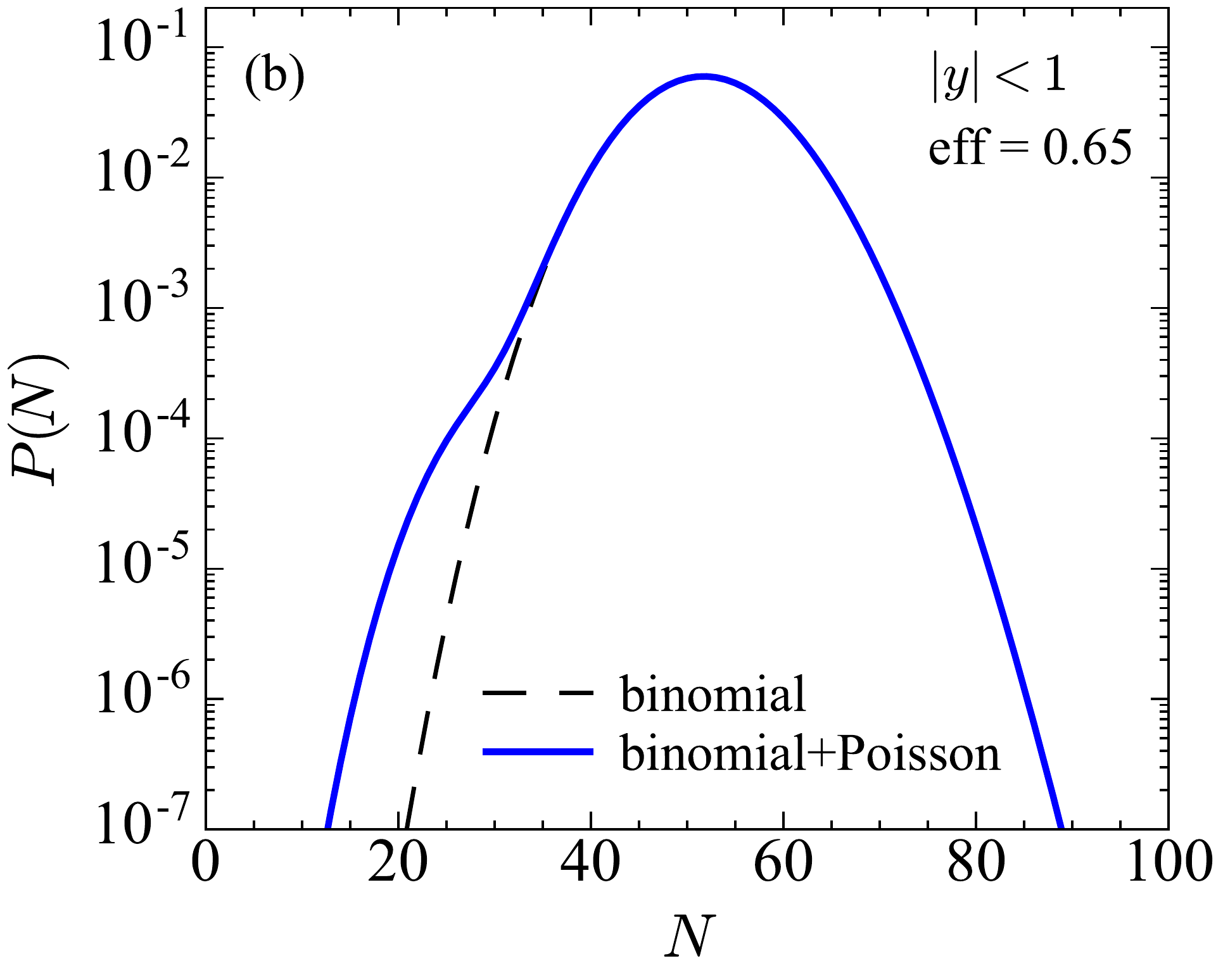}
\end{center}
%\par
%\vspace{-5mm}
\caption{$P(N)$ for 7 GeV and twice the rapidity coverage as the present preliminary STAR data. See text for details. }
\label{fig:star7_double}
\end{figure*}

For the next phase of the RHIC beam energy scan, it is expected that
STAR may be able to increase its rapidity coverage. If it could be
doubled, and the observed scaling $C_{n} \sim \ave{N}^{n}$ persists, the resulting
probability distribution would look like Fig.~\ref{fig:star7_double}.
In this case, even with an efficiency of $\epsilon\simeq 0.65$ the two
components should be visible in the efficiency uncorrected
data. 

\section{Discussion and conclusions}

To understand the relevance of the previous results several comments are in order:

%\begin{itemize}
\begin{enumerate}[(i)]
\item The factorial cumulants (integrated genuine correlation functions) 
  based on the present STAR data in central $\sqrt{s_{NN}}=7.7$ GeV collisions are consistent with the
  assumption of two distinct event classes.
  This model not only reproduces the factorial cumulants but also naturally explains the long
  correlation length observed in the STAR data. 
  
\item Provided that the measured factorial cumulants $C_n$ are much larger than expectations from an ordinary background (baryon conservation etc.) we predict that the factorial cumulants satisfy a simple relation given in Eq. (\ref{eq:ratio}), that is, $C_{n+1}/C_{n}$ does not depend on $n$. This can be tested in experiment by comparing $C_4/C_3$ with $C_5/C_4$ and $C_6/C_5$.

\item If indeed two event classes are at play, we predict
  that the 5th and 6th order factorial cumulants are very large. In addition, with
  the increasing acceptance of the STAR detector to be expected in the next phase of the RHIC
  beam energy scan, the third and forth order factorial cumulants
  should increase leading to a probability distribution which should
  exhibit a clear second event class which might even be visible
  without unfolding the data.

\item In the STAR experiment events are selected in centrality classes by the number of charged
  particles (other than protons and anti-protons) within the STAR acceptance
  \cite{Adamczyk:2013dal,Luo:2015ewa}. Therefore, in order to have two distinct event classes one of
  two things need to happen: Either there is a mechanism which removes protons from a central event
  which has many charged particles. Or, for some reason there are peripheral events, where naturally only few
  protons are stopped and brought to mid rapidity, but at the same time lots of charged particles (i.e. pions) are produced
  and the event is classified as a central one.
	The latter situation is hard to fathom (as far as originating from a physical effect). 
	However, one could imagine that the observed 
	lack of protons in some central events is compensated by an abundance of deuterons or other light nuclei.
	If true, this would result in a significant anti-correlation between the proton and deuteron number.
  
\item The present STAR dataset for $\sqrt{s_{NN}}=7.7$ GeV contains 3
  million events so that the most central 5\% correspond to 150k
  events  \cite{Adamczyk:2013dal}. Given $\alpha \approx 0.0033$ there are roughly 500 events for
  $N<20$ and it maybe worthwhile to inspect these
  events individually to see if there are some systematic deviations or
  common experimental issues. Indeed, in a recent paper \cite{He:2018mri} the possibility that a certain subset of events 
	could have different/fluctuating efficiency has been discussed, and it was shown that such a scenario
	would effectively result in two or more event classes.

\begin{figure*}[t]
\begin{center}
\includegraphics[scale=0.28]{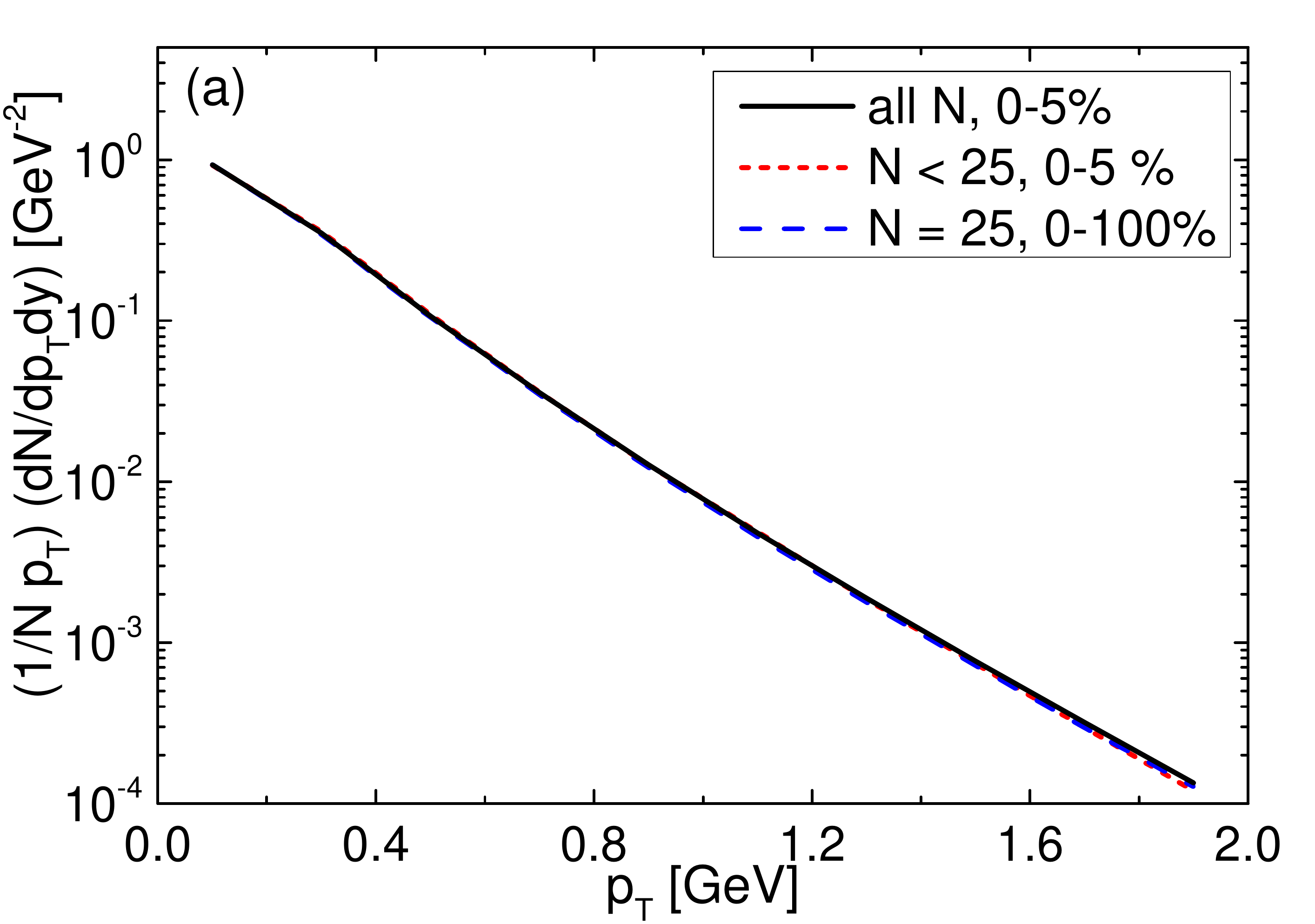} 
\includegraphics[scale=0.28]{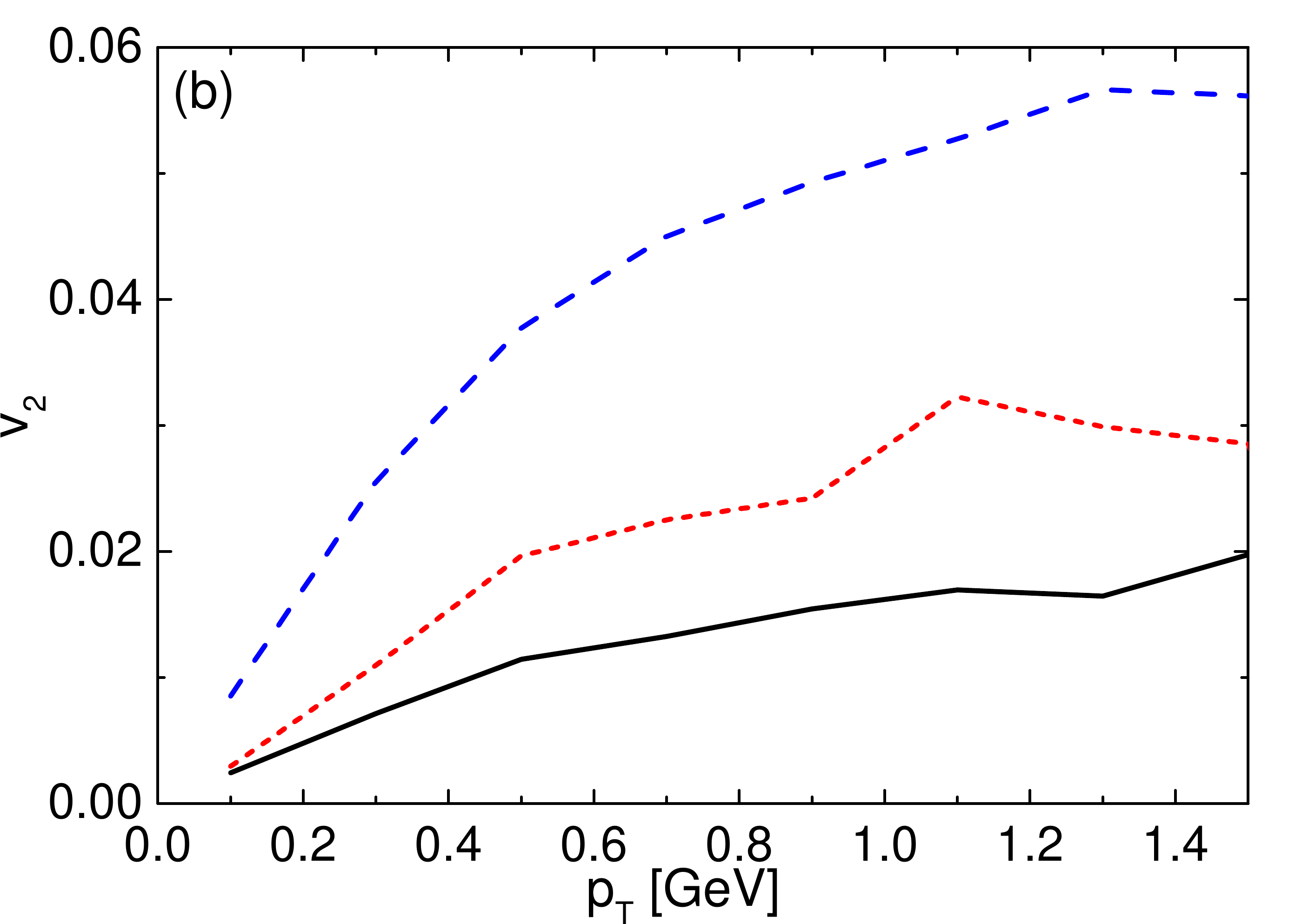} 
\end{center}
%\vspace{-5mm}
\caption{(a): Transverse momentum distributions of charged pions from the UrQMD transport model. 
(b) Elliptic flow of charged pions as a function of the transverse momentum from UrQMD simulations. In both plots three different event selections are compared. The black solid lines correspond to all most central (0-5$\%$) events, while the red short dashed lines depict only most central events with a small proton number ($N < 25$). The blue dashed lines correspond to the reference events with exactly 25 protons, but selected from all centralities.}
\label{fig_urqmd}
\end{figure*}

\begin{figure*}[t]
\begin{center}
\includegraphics[width=0.70\textwidth]{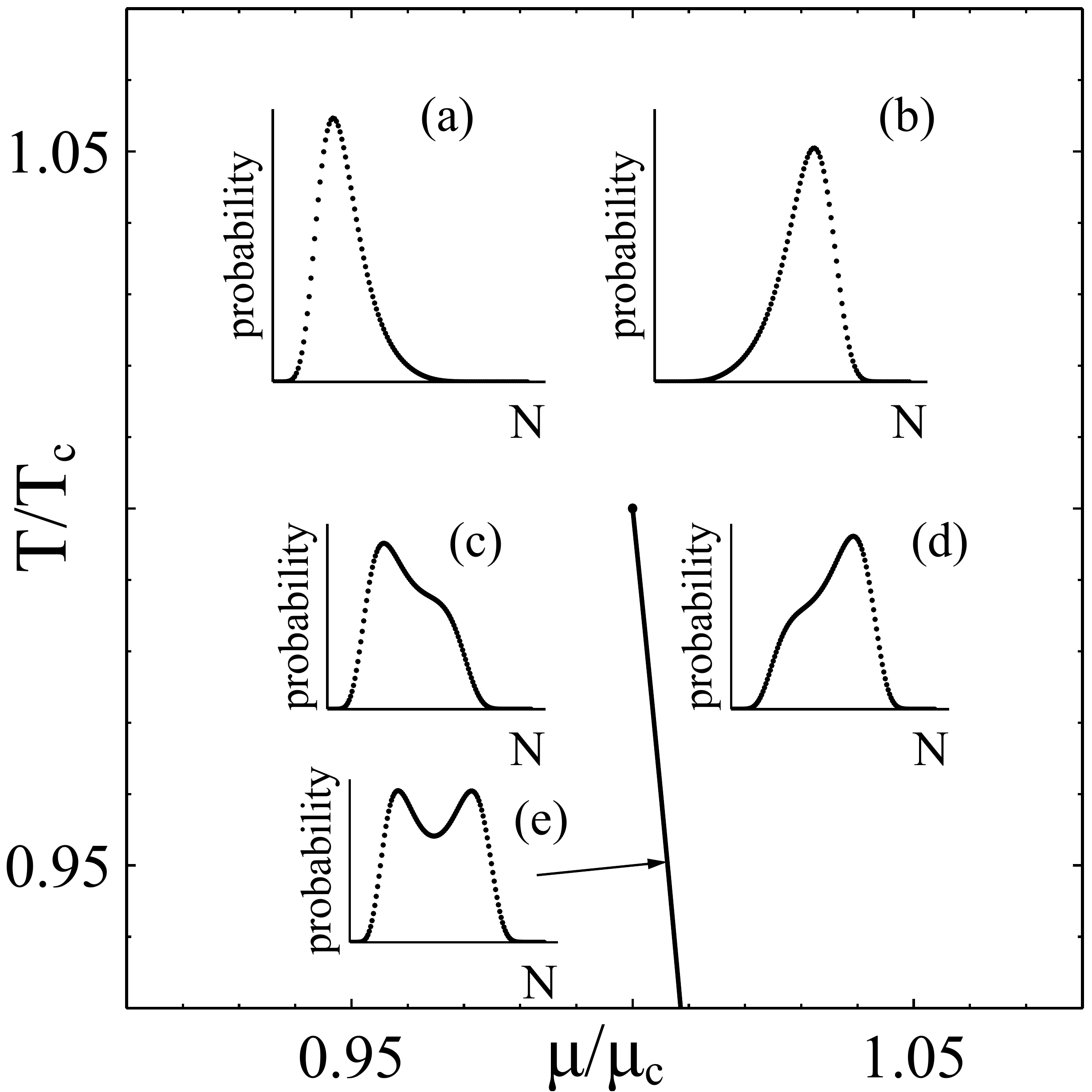}
\end{center}
%\par
%\vspace{-5mm}
\caption{Probability distribution at various points close to the
co-existence line for the van der Waals model for system at fixed volume: $(T/T_c, \mu/\mu_c) = $ (1.02, 0.99) (a), (1.02, 1.004) (b), (0.98, 1.0015) (c), (0.98, 1.004) (d), (0.95, 1.0062) (e). The model and parameters are described in the Appendix \ref{sec:app_waals}.}
\label{fig:dima_plot}
\end{figure*}

\item The natural way to explore the possible presence of distinct event classes in the STAR data,
  is to select events with a small number of protons (in central events) to enrich the contribution
  of the small event class.  Then one can investigate whether these events exhibit certain
  characteristics, which distinguishes them from all other events
  in that centrality class.  Such an event characteristic could be any other observable such as,
  e.g., the mean transverse momentum, the elliptic flow $v_2$, HBT radii etc. Of course some
  observables may change with the proton number even if there are no separate event classes. To
  illustrate this and to set some kind of baseline for the case where we do not have distinct event
  classes, we show the charged pion transverse momentum distributions and elliptic flow from UrQMD
  simulations, as function of the transverse momentum in
  Figs.~\ref{fig_urqmd} (a) and (b).  To produce these results 30 million minimum bias UrQMD events of Au+Au
  collisions at a beam energy of $\sqrt{s_{NN}}= 7.7\gev$ 
  were generated. 
	Then the 0-5$\%$ most central events where selected as the most central bin, and are plotted as
  black-full lines. Here we used the same event selection as used in the STAR analysis, namely the
  centrality was determined by the number of charged particles (except protons and anti-protons) in
  $|y|\leq 1$. The number of protons is also determined in the STAR acceptance, $|y|\leq 0.5$ 
  and $0.4 \leq p_T \leq 2.0$ GeV,
  which gives a mean net-proton number of $\left\langle N \right\rangle \approx 42$.
  From these 0-5$\%$ most central events we then selected the 0.3$\%$ events with the
  smallest number of net protons ($N < 25$), i.e., the left tail of the net proton multiplicity
  distribution, which is shown as red-dotted lines. This effect stems only from the fact that 
  even different events in a given centrality class will have varying $v_2$. 
  As a comparison we also selected, from all
  minimum bias events, only the events with a net proton number of $N = 25$, shown as
  blue-dashed lines.  Interestingly the resulting transverse momentum distributions
  seem almost independent of the event selection (the actual difference between the curves is less
  than 5\%)\\
  On the other hand, the elliptic flow, $v_2$, depends rather strongly on the event selection.
  The most central events exhibit the smallest and the more peripheral events (blue line) have the
  largest elliptic flow. However, the central events with the smallest number of protons also show
  an increased $v_2$ compared to all central events.  This is understandable, as the left tail of
  the proton distribution for the most central events contains a larger number of 'peripheral'
  events, i.e.,
  events with a larger impact parameter and therefore larger initial anisotropy.\\
  We can therefore conclude that the elliptic flow is a good criterion to select events based on
  their initial anisotropy. As a consequence, if STAR finds that the elliptic flow in the left tail (small 
  number of protons)
  of the most central events has an elliptic flow which is as large as the elliptic flow in
  peripheral events with the same proton number, the additional events are likely due to
  misidentified peripheral events. On the other hand if STAR finds that the events in the low proton
  number tail have a $v_2$ equal to or smaller than the average of all central events, interesting
  physics may be at play.  This would also be the case if one would find a significant deviation in
  the transverse momentum spectra for the evens in the lower tail of the most central events.

\item It is noteworthy that two event classes distribution looks very
  similar to that of a system close to a first order phase
  transition in a finite system. 
  To illustrate this, we have used the van der Waals model in a finite volume
  to calculate the multiplicity distributions for various points near
  the co-existence line for a system of fixed volume (details are in
  the Appendix \ref{sec:app_waals}). This is shown in Fig.~\ref{fig:dima_plot}. The multiplicity distribution extracted
  from the STAR cumulants, Fig.~\ref{fig:1}, looks qualitatively similar to the distribution to the
  right of the phase-coexistence line in  Fig.~\ref{fig:dima_plot}. In this case the ``bump'' at small $N$
  corresponds to events where the system would be in the ``dilute'' phase whereas the large maximum
  at large $N$ corresponds to the events where the system is in the ``dense'' phase, which dominates the
  distribution. 
	
\item Naturally, the knowledge of only four factorial cumulants, $C_{1}, \ldots, C_{4}$ does not uniquely
  determine the multiplicity distribution. And indeed applying the methods of \cite{Charlier_0809.4151} based
  on the Poisson-Charlier expansion (see Appendix \ref{sec:app_chalier} for some details) one can
  generate a different distribution which also reproduces the factorial cumulants of the preliminary STAR
  data.\footnote{See also Ref. \cite{Behera:2017xwg} for another way of constructing a probably distribution 
  based on the Pearson curve method.}   
  We verified that this distribution shows a
  much smaller shoulder on the left as the two event classes model used here. While the first four cumulants
  are the same for the two distributions by construction, the prediction for the fifth and sixth differ
  dramatically, as this distribution results in $C_{5}=-200$ and $C_{6}=4220$,
  which are about an order of magnitude smaller
  than that of the two event classes model. If, on the other hand, we enforce the $C_{5}$ and $C_{6}$
  to be of the same magnitude as the two classes model, the resulting probability distribution from
  the Poisson-Charlier method develops a shoulder at low $N$ similar to the one of the two classes
  model. In other words, if the magnitude of the predicted values of 
  $C_{5}$ and $C_{6}$ in Eq. (\ref{eq:C56_predict}) are
  confirmed by experiment, the presence of a shoulder at low $N$ is very likely, thus favoring the
  two event classes hypothesis.
  
\item Instead of two distinct event classes one could have the situation,
where we have two mechanisms contributing at the same time to {\em each}
event. In this case the multiplicity distribution is given by
\begin{equation}
P(N)=\sum_{N_{(a)},N_{(b)}}P_{(a)}(N_{(a)})P_{(b)}(N_{(b)})\delta
_{N_{(a)}+N_{(b)}-N},
\end{equation}
and the generating function $H(z) = H_{(a)}(z)H_{(b)}(z)$,
and the factorial cumulant generating function is $G(z) = G_{(a)}(z)+G_{(b)}(z)$.
In this case the factorial cumulants are given by%
\begin{equation}
C_{k}=C_{k}^{(a)}+C_{k}^{(b)}.
\end{equation}
Clearly, to get large $C_{k}$ we either need $C_{k}^{(a)}$ or $C_{k}^{(b)}$
to be large. This situation  would for example correspond to the case
of cluster formation discussed in \cite{Bzdak:2016jxo}, where the
presence of clusters lead to large correlations.
%\end{itemize}
\end{enumerate}

To conclude, we have considered a model where the particle (proton) multiplicity distribution arises
from two distinct event classes. We showed that in this case, the factorial cumulants of
the combined distribution can be very large, even though the distribution of the individual classes
may have small or even vanishing factorial cumulants, as in the case of Poisson distributions. We
further showed that in this picture the factorial cumulants would scale like $C_{n}\sim \ave{N}^{n}$ 
and their ratios do not depend on $n$, $\frac{C_{n+1}}{C_{n}}=const$ ($n>3$ at $7.7$ GeV). Consequently, their magnitude increases rapidly with the order of the cumulants, and the
factorial cumulants grow very fast with increasing acceptance. Both these features are seen in the presently available
preliminary STAR data, which can be reproduced in this model. Our model predicts large values for
the fifth and sixth factorial cumulants, 
and thus it can be ruled out if STAR measures values considerably smaller.

\acknowledgments We thank A. Bia{\l}as for helpful discussions.  A.B. is partially supported by the
Faculty of Physics and Applied Computer Science AGH UST statutory tasks No. 11.11.220.01/1 within
subsidy of Ministry of Science and Higher Education, and by the National Science Centre, Grant
No. DEC-2014/15/B/ST2/00175.  V.K. and D.O. were supported by the U.S. Department of Energy, Office
of Science, Office of Nuclear Physics, under contract number DE-AC02-05CH11231.  D.O. also
received support within the framework of the Beam Energy Scan Theory (BEST) Topical Collaboration.
The computational resources where in part provided by the LOEWE Frankfurt Center for Scientific
Computing (LOEWE-CSC). JS thanks the Alexander von Humboldt Stiftung for supporting his stay at LBNL.

\appendix

\section{Results for $C_{5}$ and $C_{6}$}
\label{sec:app_C5_and_C6}
Here we list the result for the factorial cumulants $C_{5}$ and $C_6$
in the two component model, Eq. \eqref{eq:two_component}. First the
full result:
\begin{eqnarray}
C_{5} = C_{5}^{(a)} - &&
     \alpha  \left\{ \dc{5} +
\right.
\non 
       (1-\alpha ) && \left[ (1-2 \alpha ) \left(1-12 \alpha +12
         \alpha^2\right) \nb^5
       -10 \left(1-6 \alpha +6 \alpha ^2\right) \nb^3 \dc{2}
\right.
\non 
       && \left. \left.
+10 (1-2 \alpha )
   \nb^2 \dc{3}+15 (1-2 \alpha ) \nb (\dc{2})^2 -5 \nb\dc{4}
     -10 \dc{2} \dc{3}
       \right]
  \right\} ,
\non
C_{6}=  C_{6}^{(a)}  -  &&\alpha \left\{ \dc{6} -
\right.
 \non
  (1-\alpha ) && 
\left[   
\left(1-30 \alpha  (1-\alpha ) (1-2 \alpha )^2\right) \nb^6 
-15 (1-2 \alpha ) \left(1-12 \alpha +12 \alpha ^2\right) \nb^4 \dc{2}
     \right.
\non
     &&
+20 \left(1-6
   \alpha +6 \alpha ^2\right) \nb^3 \dc{3}
-15 \nb^2 
\left(\dc{4} (1-2 \alpha )-3 (\dc{2})^{2}
 \left(1-6 \alpha +6 \alpha ^2\right)\right)
\non
     && 
+6 \nb \left( \dc{5}-10 \dc{2} \dc{3} (1-2 \alpha
   )\right)
-15 (1-2 \alpha ) (\dc{2})^{3} 
\non
   && \left. \left.
+ 10 (\dc{3})^{2}+15 \dc{2}
        \dc{4}
   \right]
\right\} .
  \label{}
\end{eqnarray}

Next the case  of small or vanishing factorial cummulants of the
individual distributions, $C_{n}^{(a)}\simeq C_{n}^{{b}}\simeq 0$
\begin{eqnarray}
C_{5} &=&C_{5}^{(a)}-\alpha (1-\alpha) (1-2\alpha )(1-12\alpha
+12\alpha^{2}) \nb^{5} \approx C_{5}^{(a)}-\alpha \nb^{5},
\notag \\
C_{6} &=&C_{6}^{(a)}+\alpha (1-\alpha) (1-30\alpha (1-\alpha
)(1-2\alpha )^{2}) \nb^{6} \approx C_{6}^{(a)}+\alpha \nb^{6},
\end{eqnarray}
where on the RHS we assumed $\alpha \ll 1$.

\section{Multiplicity distribution in a finite-volume van-der Waals model}
\label{sec:app_waals}

Here we illustrate, how a multiplicity distribution visually similar to Figs.~\ref{fig:1} and \ref{fig:star7_double}
emerges in a simple toy model of interacting nucleons in a finite
volume $V$. To this end we use the well known van-der-Waals model (see,
e.g., Ref. \cite{reif_book}) which combines a
repulsive interaction, realized via an excluded volume $b$, and an attractive
interaction by means of the 
mean-field potential, which is proportional to density: $U = a \frac{N}{V}$. The canonical partition function 
of this model reads
\begin{eqnarray}
 \mathcal{Z}_{CE}(N, V, T) &=&  \frac{1}{N!} \int \prod \frac{d^3 x_i d^3p_i}{(2\pi\hbar)^{3 N}} e^{- U N/T} \notag \\
                  &=&  \frac{1}{N!} \left(V \varphi(T) \right)^N \left(1 - \frac{b N}{V} \right)^N e^{\frac{a N^2}{V T}} \,,
\end{eqnarray}
where
$\varphi(T) = \frac{g}{2\pi^2} T^3 \left(\frac{m}{T}\right)^2
K_2\left(\frac{m}{T}\right)$, $g=4$ being the degeneracy of the
nucleons and $m$ the proton mass. If the volume $V$ is embedded into a
thermal bath with baryochemical potential $\mu$, the probability $w_N$
to find $N$ protons inside of the volume is given by
\begin{eqnarray} \label{eq:vdw_probability_distribution}
 w_N = \frac{e^{\mu N /T} \mathcal{Z}_{CE}(N, V, T)}{\sum_{N} e^{\mu N /T} \mathcal{Z}_{CE}(N, V, T)} .
\end{eqnarray}
It is this probability which is plotted in Fig.~\ref{fig:dima_plot}.

The denominator of Eq.~\eqref{eq:vdw_probability_distribution} is the
grand-canonical partition function,\\
$\Omega (\mu, V, T) = \sum_{N} e^{\mu N /T} \mathcal{Z}_{CE}(N, V,
T)$. To verify, that the above expression for the probability $w_{N}$
is really that of the van-der-Waals model, let us consider the
thermodynamic limit ($V \to \infty$, $N \to \infty$, $N/V \to n$) and
extract the resulting equation of state. In this case the
grand-canonical partition function $\Omega$ can be approximated by the
largest term of the sum, determined by
\begin{eqnarray}
 \frac{\partial \log \left(e^{\mu N /T} \mathcal{Z}_{CE}(N, V, T) \right)}{\partial N} \bigg\rvert_{N = N^*} = 0 \,.
\end{eqnarray}
This condition is equivalent to
\begin{eqnarray}
 \frac{N^*}{\varphi(T) (V - b N^*)} = \exp \left( \frac{\mu}{T} - \frac{b N^*}{V - b N^*} +
  \frac{2 a N^*}{V T} \right) \, ,
\end{eqnarray}
which coincides with the grand-canonical van-der-Waals equation of state \cite{Vovchenko:2015xja}. Indeed, substituting it
to the expression for $\Omega$ and using $p V = T \log \Omega$, one obtains familiar $(p + a n^2)(1 - n b) = n T$, where $n = N^*/V$.

For reference, the parameters $a$ and $b$ are chosen such that the critical temperature
$T_c = \frac{9 a}{27 b} = 0.070$ GeV and critical density $\rho_c = \frac{1}{3b} = 0.3$
GeV/fm$^3$. The volume is taken to be $V = 200$ fm$^3$.

\section{Constructing a discreet distribution from given factorial cumulants}
\label{sec:app_chalier}

In the main body of this paper the first four factorial cumulants $C_{1-4}$ of the proton
multiplicity distribution are described based on a specific form for the distribution. Of course, the
first four factorial cumulants do not uniquely determine the distribution. It is therefore
useful to explore, how different  discreet distributions with the same set of factorial cumulants could be.
To find out, here we provide a simple way to construct a (non-unique) distribution which will return
a given set of factorial cumulants, $C_{1}\ldots C_{n}$, inspired by the Poisson-Charlier expansion \cite{Charlier_0809.4151}.

Let us denote the Poisson distribution with mean $\mu$ by
$\pi_{\mu}(k) = \frac{\mu^k}{k!}e^{-\mu}$ and assume that $\pi_{\mu}(k < 0) = 0$.
Let us also introduce a forward difference operator $\nabla$ acting on an
arbitrary discreet function $\varphi$:
\begin{eqnarray}
\nabla \varphi(k) &=& \varphi(k) - \varphi(k - 1) \\
\nabla^{l} \varphi(k) &=& \sum_{j = 0}^l \binom{l}{j} (-1)^j \varphi(k - j) \,.
\end{eqnarray}

Next one introduces the distribution  $\mathit{f}_n$ which depends on the first $n$ factorial
cumulants, $C_{1}\ldots C_{n}$ as follows: First one defines an operator $D_{n}(t)$ which represents all the terms of the
Taylor expansion in $t$ of 
\begin{eqnarray}
\exp \left(\sum_{j=2}^{\infty} \frac{C_j}{j!} (-t \nabla)^j \right) = D_{n}(t) + {\cal O}(t^{n+1})
\end{eqnarray}
up to order $t^{n}$.  This ensures that $D_{n}(t)$ involves no factorial cumulants of order
$k>n$. Given the operator $D_{n}(t)$ the distribution $f_{n}(k)$ is then defined as 
\begin{eqnarray}
f_{n}(k) = D_{n}(t=0) \, \pi_{\mu}(k) \,.
\end{eqnarray}
For example, for the first four orders, this results in
\begin{eqnarray}
\mathit{f}_1(k) &=& \pi_{\mu}(k) \\
\mathit{f}_2(k) &=& \pi_{\mu}(k) + \frac{C_2}{2}  \nabla^2 \pi_{\mu}(k) \\
\mathit{f}_3(k) &=& \pi_{\mu}(k) + \frac{C_2}{2}  \nabla^2 \pi_{\mu}(k) - \frac{C_3}{6}  \nabla^3 \pi_{\mu}(k)\\
\mathit{f}_4(k) &=& \pi_{\mu}(k) + \frac{C_2}{2}  \nabla^2 \pi_{\mu}(k) - \frac{C_3}{6}  \nabla^3 \pi_{\mu}(k) +
                    \frac{3C_2^2 + C_4}{24} \nabla^4 \pi_{\mu}(k).
\label{eq:fk_charlier}
\end{eqnarray}

The functions $\mathit{f}_n$ are normalized by construction. Indeed, it is easy to see
that $\sum_{k} \nabla^{l} \pi_{\mu}(k) = \sum_{j = 0}^l \binom{l}{j} (-1)^j = (1-1)^l = 0$ for any $l > 0$. Therefore
$\sum_{k} \mathit{f}_n (k) = \sum_{k} \pi_{\mu}(k) = 1$. 

The higher factorial cumulants of $\mathit{f}_n$ are not zero, as one might naively expect. For
example, in case of $\mathit{f}_4$ one gets
\begin{eqnarray}
  C_2^{\mathit{f}_4} &=& C_2 \\
  C_3^{\mathit{f}_4} &=& C_3 \\
  C_4^{\mathit{f}_4} &=& C_4 \\
  C_5^{\mathit{f}_4} &=& -10 C_2 C_3 \\
  C_6^{\mathit{f}_4} &=& -5 \left( 3 C_2^3 + 2 C_3^2 + 3 C_2 C_4 \right) \,.
\end{eqnarray}

Finally let us prove, that the first $n$ factorial cumulants of $f_n$
are indeed $C_1$, $C_2$, $\dots$, $C_n$. The proof is by induction. We first show that
the first $n$ factorial cumulants$\mathit{f}_{n+1}$ and
$\mathit{f}_{n}$ are identical. Then we use the result of Appendix B from
\cite{Charlier_0809.4151}, that all the factorial cumulants
of $\mathit{f}_{\infty}$ coincide with $C_i$ ($i = 1 \dots \infty$) used to construct it.
Let $F_n(z)$ be factorial cumulant generating function of $\mathit{f}_n$,
\begin{eqnarray}
F_n(z) = \log \left( \sum_{k = 0}^{\infty}\mathit{f}_n(k) (z)^k \right) \,.
\end{eqnarray}

Then $F_{n+1}$ and $F_n$ are connected by
\begin{eqnarray}
e^{F_{n+1}(z)} &=& e^{F_n(z)} + A \sum_{k = 0}^{\infty} \nabla^{n+1} \pi_{\mu}(k) (z)^k  \notag \\
               &=& e^{F_n(z)} + A (1-z)^{n+1} e^{\mu (z-1)}\,,
\end{eqnarray}
where $A$ is some expression containing the factorial cumulants, $C_{1}\ldots C_{n+1}$, (see
e.g. Eq.\eqref{eq:fk_charlier}) and thus is a real number independent on $z$ and $k$. 
Then
\begin{eqnarray}
F_{n+1}(z) = F_n(z) + \log \left( 1 + A (1-z)^{n+1} e^{\mu (z-1) - F_n(z)} \right) \,,
\end{eqnarray}
and one can see that the Taylor expansion of the second term in $z$ around $z=1$ starts with the
power $n+1$, since $F_n(z)$ is a polynomial in $z-1$. Therefore,
the first $n$ factorial cumulants of $\mathit{f}_{n+1}$ and $\mathit{f}_n$ are identical. In other words, adding terms
of order larger than $n$ does not change the first $n$ factorial cumulants. Hence, since $\mathit{f}_{\infty}$ and $\mathit{f}_n$ coincide
up to order $n$, their first $n$ factorial cumulants are identical and equal to $C_1$, $\dots$, $C_n$.

Unfortunately, $f_n(k)$ is not always a proper probability mass function, because it can become
negative. So one needs to explicitly verify if a given expansion is actually positive definite. The
resulting distribution based on the STAR factorial cumulants, Eq.~\eqref{eq:STAR_fact_cumulants}, is
indeed positive definite and, of course, normalized as proven above. 

\bibliography{paper}
%\bibliography{/home/dima/Work/Fluctuations_project/paper/Jan_bug/Jan_bug/paper}
%\bibliography{/Users/vkoch/Documents/Bibliography/myBibliography,/Users/vkoch/Documents/Bibliography/myPapers}
\end{document}

%\preprint{}
\title{
Addendum to "Large proton cumulants from the  superposition of ordinary multiplicity distributions"
}

\author{Adam Bzdak}
%\email{bzdak@fis.agh.edu.pl}
\affiliation{AGH University of Science and Technology,
Faculty of Physics and Applied Computer Science,
30-059 Krak\'ow, Poland}

\author{Volker Koch}
%\email{vkoch@lbl.gov}
\affiliation{Nuclear Science Division, 
Lawrence Berkeley National Laboratory, 
Berkeley, CA, 94720, USA}

\author{Dmytro Oliinychenko}
%\email{DOliinychenko@lbl.gov}
\affiliation{Nuclear Science Division, 
Lawrence Berkeley National Laboratory, 
Berkeley, CA, 94720, USA}

\author{Jan Steinheimer}
%\email{steinheimer@fias.uni-frankfurt.de}
\affiliation{Frankfurt Institute for Advanced Studies, 
Ruth-Moufang-Str. 1, D-60438 Frankfurt am Main, Germany}

\begin{abstract}
We show that the large values of proton factorial cumulants $C_{5} \approx -2645 $ and $C_{6}
\approx  40900$ at $\sqrt{s}=7.7$ GeV, predicted in our recent publication \cite{Bzdak:2018uhv}, can
in principle be verified or ruled out  with the current STAR statistics of approximately $150000$
events in the $0-5\%$ centrality class.
\end{abstract}

\maketitle

%\bigskip

In our recent paper \cite{Bzdak:2018uhv} we argued that the surprisingly large proton cumulants (and
extracted factorial cumulants) measured by the STAR collaboration at $\sqrt{s}=7.7$ GeV in $0-5\%$
central Au+Au collisions \cite{Luo:2015ewa,Bzdak:2016sxg} can be understood in a two component model
\begin{equation}
P(N)=(1-\alpha )P_{(a)}(N)+\alpha P_{(b)}(N),
\label{eq:two_component}
\end{equation}
where $\alpha \ll 1$ and both $P_{(a)}$  
and $P_{(b)}$ are the proton multiplicity distributions characterized by small or even vanishing
factorial cumulants. As discussed in \cite{Bzdak:2018uhv}, such a distribution would be consistent
with the system being close to a first order phase transition.

In \cite{Bzdak:2018uhv} we argued that Eq. (\ref{eq:two_component}) can naturally describe the 
STAR preliminary data up to
the fourth-order cumulant (and factorial cumulant). The higher order factorial cumulants,
predicted in \cite{Bzdak:2018uhv}, are  given by 
\begin{equation}
C_{5} \approx -2645, \quad\quad C_{6} \approx  40900.
  \label{eq:C_predict}
\end{equation}  

In this note we show that this result can be verified (or ruled out) already with the current STAR statistics of
144393 events\footnote{We thank Xiaofeng Luo for providing this number.} in $0-5\%$ centrality
class. While the number of events is rather limited, as we will show, the statistics is sufficient
to test the predictions of our model. 

\begin{figure}[t]
\begin{center}
\includegraphics[scale=0.35]{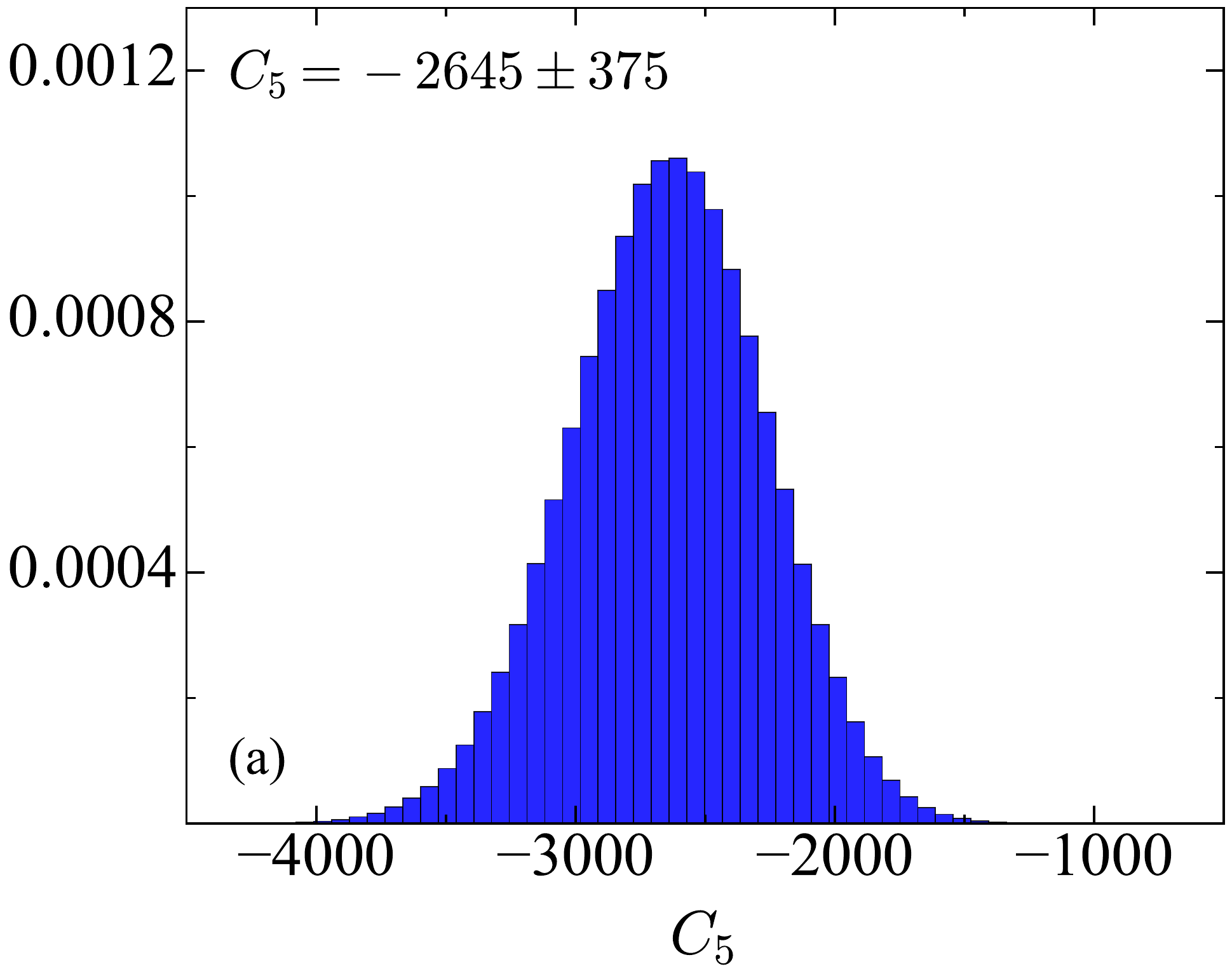} \hspace{5mm}
\includegraphics[scale=0.35]{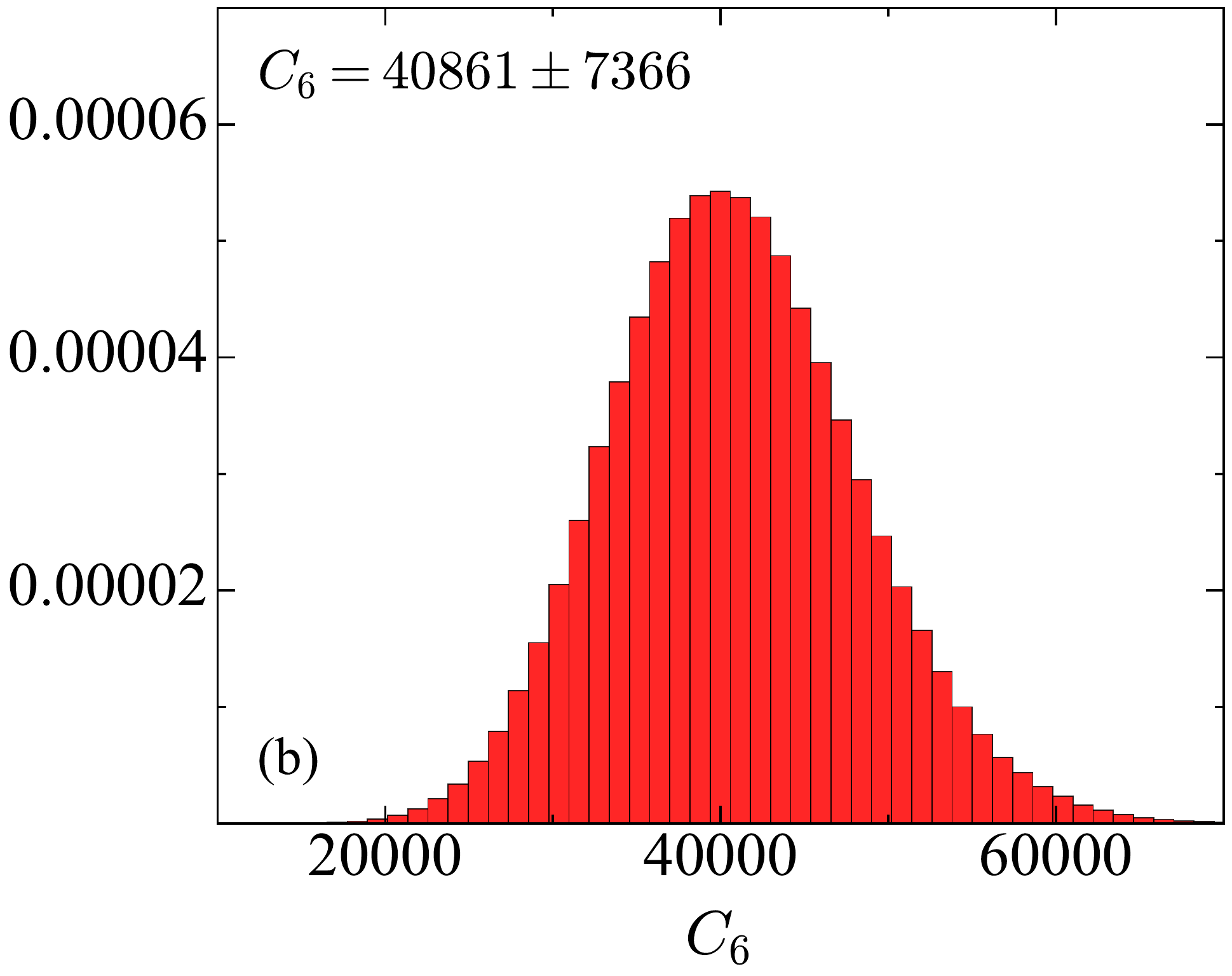}
\end{center}
\par
\vspace{-5mm}
\caption{Histograms (normalized to unity) of (a) factorial cumulant $C_{5}$ and (b) factorial cumulant $C_{6}$, based on 144393 events sampled from the distribution given by Eq. (\ref{eq:two_component}).}
\label{fig:1}
\end{figure}

In Fig. \ref{fig:1} we show our results based on the current STAR statistics of 144393 events for
the factorial cumulants, $C_{5}$, and $C_{6}$. In 
our exercise we sample the number of protons, $N$, $144393$ times from the distribution $P(N)$
given by Eq. (\ref{eq:two_component}). Then we calculate  the values of $C_{n}$, $n=4,5,6$. Finally
the previous steps are repeated a sufficient number of times to obtain a smooth histogram of $C_{n}$
values, which provides an estimator of the statistical error. We find that $C_{4}$ agrees with
the value of $C_{4}\simeq 170$ \cite{Bzdak:2016sxg} extracted from the STAR data, which has to be
expected since the parameters for the distribution, Eq. \eqref{eq:two_component}, where adjusted to do just
that. More importantly, the 
results for $C_{5}$ and $C_{6}$ shown in the figure, suggest that the prediction from our paper,
$C_{6}\gg-C_{5}\gg C_{4}$, or equivalently $C_{6}/C_{5}\simeq C_{5}/C_{4}\simeq {\cal O}(10)$ can be either verified or
ruled out given the statistics of the presently available STAR data set.

%\acknowledgments 
\bigskip
\bigskip
{\bf Acknowledgments:} We thank Xiaofeng Luo for helpful correspondence. A.B. is partially supported by the
Faculty of Physics and Applied Computer Science AGH UST statutory tasks No. 11.11.220.01/1 within
subsidy of Ministry of Science and Higher Education, and by the National Science Centre, Grant
No. DEC-2014/15/B/ST2/00175. V.K. and D.O. were supported by the U.S. Department of Energy, Office
of Science, Office of Nuclear Physics, under contract number DE-AC02-05CH11231.  D.O. also
received support within the framework of the Beam Energy Scan Theory (BEST) Topical Collaboration.
The computational resources where in part provided by the LOEWE Frankfurt Center for Scientific
Computing (LOEWE-CSC). JS thanks the Alexander von Humboldt Stiftung for supporting his stay at LBNL. 

\bibliography{paper}